\documentstyle[amsmath,amssymb,amsfonts,graphicx,epsfig,12pt]{elsart}
\begin{document}
\begin{frontmatter}
\title{Tensor gauge field localization on a string-like defect}
\author[UFC,CEFET1]{L. J. S. Sousa},
\author[CEFET2]{W. T. Cruz}, and
\author[UFC]{C. A. S. Almeida}
\address[UFC]{ Departamento de
F\'{\i}sica - Universidade Federal do Cear\'{a} \\ C.P. 6030, 60455-760 Fortaleza-Cear\'
{a}-Brazil}
\address[CEFET1]{Instituto Federal de Educa\c{c}\~{a}o, Ci\^{e}ncia e Tecnologia do Cear\'{a} \\ Campus de Canind\'{e}, 62700-000
Canind\'{e}-Cear\'{a}-Brazil}
\address[CEFET2]{Instituto Federal de Educa\c{c}\~{a}o, Ci\^{e}ncia e Tecnologia do Cear\'{a} (IFCE), Campus Juazeiro do Norte, 63040-000 Juazeiro do Norte-Cear\'{a}-Brazil}
%\thanks[e-mail]{Electronic addresses: wilami,carlos@fisica.ufc.br}
\begin{abstract}
This work is devoted to the study of tensor gauge fields on a string-like defect in six dimensions. This model is very successful in localizing fields of various spins only by gravitational interaction. Due to problems of field localization in membrane models we are motivated to investigate if a string-like defect localizes the Kalb-Ramond field. In contrast to what happens in Randall-Sundrum and thick brane scenarios we find a localized zero mode without the addition of other fields in the bulk. Considering the local string defect we obtain analytical solutions for the massive modes. Also, we take the equations of motion in a supersymmetric quantum mechanics scenario in order to analyze the massive modes. The influence of the mass as well as the angular quantum number in the solutions is described. An additional analysis on the massive modes is performed by the Kaluza-Klein decomposition, which provides new details about the KK masses.
\end{abstract}
\end{frontmatter}

%==================================================================================

\section{Introduction}

The idea that our world is a three brane embedded in a higher-dimensional space-time has attracted
the attention of the physics community basically because this give solution for some intriguing problems
in the Standard Model, like the hierarchy problem, the dark matter origin and the cosmological constant problem \cite{Rubakov1983, YuriShtanov2009, G.D.Starkman2001a, G.D.Starkman2001}. There are, at least, two kinds of theories that carry this basic idea, namely that proposed by Arkani-Hamed, Dimopoulos and Dvali \cite{I.Antoniadis1998, N.Arkani-Hamed2000} and the so called Randall - Sundrum models \cite{Randall1999a,Randall1999}. In these models it is assumed that, in principle, all the matter fields are constrained to propagate only on the brane, whereas gravity is free to propagate in the extra dimensions. Others models are considered, where the bulk is endowed not only with gravity, but torsion too \cite{MukhopadhyayaB2002, Mukhopadhyaya2004, Lebedev2002}. For the 4D case the question if torsion is present or not is only of academical interest, because its effects are enormously suppressed. In space-time with extra dimensions however the situation changes and the inclusion of torsion may be of great interest \cite{Lebedev2002}, which makes more appealing that kind of models.

Apart from the gravity and torsion fields all the other fields of the Standard Model are, a priori, assumed
to be trapped on the brane, in the models considered above. But this assumption is not so obvious, therefore it is
interesting to look for alternative field theoretic localization mechanism \cite{Oda2000}. In this way, many
works have been published in the last decade, concerning to localization mechanism in the Randall-Sundrum model,
in $AdS_{5}$ space \cite{Mukhopadhyaya2004, Davoudias2000, Kehagias2001, Tahim2009}. It is well known that,
particularly, spin-0 and graviton fields are localized on the brane but this is not the case for the spin-1
field. The Kalb-Ramond tensor field possesses a localized zero mode strongly suppressed by the size of the extra
dimension \cite{Mukhopadhyaya2004}. In reference \cite{Tahim2009} the authors showed that the Kalb-Ramond zero
mode localization is possible only with the presence of the dilaton field through a coupling to the Kalb-Ramond
field.

Once we assume that our world may be a 3-brane or a 4-brane on a six dimensional space-time, it is natural to
look for mechanism of field localization in this kind of geometry. In addition, we are motivated to investigate
if a string-like defect localizes fields in a more 'natural' way than the domain wall geometry. As a matter of fact,
we know that the gravitational interaction is not sufficient to localize spin 1 field and Kalb-Ramond field.
For instance, in the Randall-Sundrum model \cite{Davoudias2000,Tahim2009} an additional interaction of a dilaton field is necessary. Therefore, a string-like defect may be a more convenient model for our universe if it is
possible to localize the Standard Model fields in this kind of geometry only through gravitational interaction.

Several works have been done in this subject and it has been shown that indeed most of the Standard Model
 fields are localized on a string-like defect. We know that spin 0, 1, 2, 1/2 and 3/2 fields are all localized
on a string-like scenario. The bosonic fields are localized with exponentially decreasing warp factor and the fermionic fields
 are localized on defect with increasing warp factor \cite{Oda2000a, Gherghetta2000}. What is interesting here is that spin 1 vector field that is not localized on a domain wall in Randall-Sundrum model, can be localized in the string-like defect. This fact encourages us to look for localization of the Kalb-Ramond field on the string-like defect.

The mathematical way to verify if a field is localized on a brane, may be summarized as follows: for models with one non compact extra dimension, say $r$, in order to find localized zero modes on the brane we must analyze the integral on this extra dimension or, in other words, to investigate the normalizability of the ground state wave function associated with it. Finite or infinite values for the integral imply in localization or non localization, respectively. The massive modes are evaluated by performing the so called Kaluza-Klein decomposition \cite{Oda2000, Oda2000a}. Another approach to treat the massive modes is to cast the equation for the $r$ in a Schroedinger-like equation and then solve it. In many works this resulting equation is not analytically solved. So in order to analyze the possibility that massive modes could be localized on the brane the resonant modes must be searched.

Here we aim to analyze the mechanisms of field localization for the Kalb-Ramond tensor field on a string-like defect, $AdS_{6}$ space. As the case of the gauge field that is localized on this geometry by means of the gravity interaction only, we hope that the Kalb-Ramond field  presents the same behavior.  We will use both the mechanisms discussed above to deal with the massive modes. As we will see, in our case the Schrodinger-like equation is analytically solvable and we can compare these two methods of analyzing the massive modes.

This work is organized as follows: in the section (2) we present a briefly review of the string-like defect; in section (3) we discuss about the field localization procedures in this kind of geometry. The zero mode and massive modes localization of the KR field are studied in subsections (3.1) and (3.2) respectively. In subsection 3.3 we analyze the massive modes via KK decomposition. Finally, in section (4) we present a discussion of our results.

\section{String-like Defect: a briefly review}

The first works on brane world in six dimensions date back to the 80's. In 1982 Akama \cite{Akama2000} used the dynamics of the Nielsen-Olesen vortex to localize our world in a three brane, embedded in a six dimensional space-time. Next, similar works were done in 1983 by Rubakov and Shaposhnikov \cite{Rubakov1983} and in 1985 by Visser \cite{Visser1985}. The emergence of the Arkani-Hamed-Dimopoulos-Dvali model \cite{I.Antoniadis1998, N.Arkani-Hamed2000} and Randall-Sundrum model \cite{Randall1999a,Randall1999} give us other classes of alternatives to the usual Kaluza-Klein compactification.

The work of Randall-Sundrum was extended to more than five dimension as soon as it appeared in literature. Particularly, in six dimensions, we can cite some authors that contributed in a space of less than a year after the original Randall-Sundrum work appearance, namely Gregory, Rubakov and Sibiryakov \cite{Gregory2000a}; Gherghetta and Shaposhnikov \cite{Gherghetta2000}; and Oda \cite{Oda2000}.

We now summarize the string-like solution of the Einstein equations. We will follow the approaches by Oda \cite{Oda2000, Oda2000a}, where more details can be seen.

We begin by considering the general metric ansatz in D-dimensional space-time
\begin{eqnarray}\label{metric}
\lefteqn {ds^{2}=g_{MN}dx^{M}dx^{N}}\nonumber\\
& &=g_{\mu\nu}dx^{\mu}dx^{\nu}+\tilde{g_{ab}}dx^{a}dx^{b}\nonumber\\
& &=e^{-A(r)}\hat{g}_{\mu\nu}dx^{\mu}dx^{\nu}+dr^{2}+e^{-B(r)}d\Omega_{n-1}^{2},
\end{eqnarray}
where $M, N,...$ denote D-dimensional space-time indices, $\mu, \nu, ...$ p-dimensional brane ones (we assume $p \geq 4$), and $a, b, ...$ denote n-extra
spatial dimension ones.

The action we assume in this work is given by
\begin{equation}\label{action}
S=\frac{1}{2\kappa_{D}^{2}} \int {d^{D}x \sqrt{-g}(R - 2\Lambda)} + \int {d^{D}x \sqrt{-g} L_{m}},
\end{equation}
where $\kappa _{D}$ is the D-dimensional gravitational constant, $\Lambda$ is the bulk cosmological constant and $L_{m}$ is some matter field Lagrangian.

The Einstein equations are obtained by variation of the action (\ref{action}) with respect to the D-dimensional metric tensor $g_{MN}$
\begin{equation}\label{einstein}
R_{MN}-\frac{1}{2} g_{MN}R = -\Lambda g_{MN} + \kappa_{D}^{2}T_{MN}.
\end{equation}

We adopt, for the energy-momentum tensor, $T_{MN}$, the following ansatz,
\begin{equation}\label{tensor}
T_{\nu} ^{\mu}=\delta _{\nu} ^{\mu}t_{0}(r),
T_{r} ^{r} = t_{r}(r),
T_{\theta_{2} } ^{\theta_{2}} = T_{\theta_{3} } ^{\theta_{3}} = ... = T_{\theta_{n} } ^{\theta_{n}} =t_{\theta}(r),
\end{equation}
where $t_{i} (i = o, r, \theta)$ are function only of $r$, the radial coordinate. With this ansatz to the energy-momentum tensor we keep spherical symmetry.

The next step is to use the ansatzs (\ref{metric}) and (\ref{tensor}) to rewrite the Einstein equations (\ref{einstein}). A straightforward calculation give us the following results
\begin{eqnarray}\label{eq1}
\lefteqn {e^{A(r)} \hat{R} - \frac{p(n - 1)}{2} A'(r) B'(r) - \frac{p(p - 1)}{4} (A'(r))^{2} + }\nonumber\\
& & - \frac{(n - 1)(n - 2)}{4} (B'(r))^{2} + (n - 1)(n - 2) e^{B(r)} - 2 \Lambda + 2 \kappa_{D} ^{2} t_{r} = 0,
\end{eqnarray}
\begin{eqnarray}\label{eq2}
\lefteqn {e^{A(r)} \hat{R} + (n - 2) B''(r) - \frac {p(n - 2)}{2} A'(r) B'(r)} \nonumber\\
& & - \frac{p(p + 1)}{4} (A'(r))^{2}- \frac{(n - 1)(n - 2)}{4} (B'(r))^{2}  \nonumber\\
& & + (n - 2)(n - 3) e^{B(r)} + pA''(r)- 2 \Lambda + 2 \kappa_{D} ^{2} t_{\theta} = 0,
\end{eqnarray}
\begin{eqnarray}\label{eq3}
\lefteqn {\frac{p - 2}{p} e^{A(r)} \hat{R} + (p - 1) \left( A''(r) - \frac{n - 1}{2} A'(r) B'(r)\right)} \nonumber\\
& & - \frac{p(p - 1)}{4} (A'(r))^{2} + \nonumber\\
& & + (n - 1)\left[B''(r) - \frac{n (B'(r))^{2}}{4} + (n - 2) e^{B(r)} \right] \nonumber\\
& &  - 2 \Lambda + 2 \kappa_{D} ^{2} t_{0} = 0.
\end{eqnarray}
Another equation is provided by the energy-momentum tensor conservation law
\begin{equation}\label{eq4}
t'_{r} = \frac{p}{2}A'(r) (t_{r}(r) - t_{0}(r)) + \frac{n -1}{2} B'(r)(t_{r}(r) - t_{\theta}(r)),
\end{equation}
where prime denotes differentiation with respect to $r$. The scalar curvature $\hat{R}$ associated with the brane metric $\hat{g}_{\mu \nu}$ and the brane
cosmological constant $\Lambda_{p}$ are related by
\begin{equation}\label{cosmo}
\hat{R}_{\mu\nu} - \frac{1}{2} \hat{g}_{\mu \nu} = -  \Lambda_{p}  \hat{g}_{\mu \nu}.
\end{equation}

The Einstein equations and the conservation law for the energy-momentum tensor will give us the possibility to calculate the functions $A(r)$ and $B(r)$. It is possible to derive this functions from the dynamics of scalar fields \cite{Koley2007}, for example. But here we will assume the ansatz $A(r) = cr$, where $c$ is a constant, for the warp factor. Now restricting ourselves to six dimensions and setting $n = 2$, the equations (\ref{eq1}), (\ref{eq2}), (\ref{eq3}) and (\ref{eq4}) assume the respective simpler forms
\begin{equation}
e^{cr} \hat{R} - \frac{p}{2} c B'(r) - \frac{p(p - 1)}{4} (c)^{2} +
- 2 \Lambda + 2 \kappa_{D} ^{2} t_{r} = 0,
\end{equation}
\begin{equation}
e^{cr} \hat{R}  - \frac{p(p + 1)}{4} (c)^{2} - 2 \Lambda + 2 \kappa_{D} ^{2} t_{\theta} = 0,
\end{equation}
\begin{eqnarray}\label{eq5}
\lefteqn {\frac{p - 2}{p} e^{cr} \hat{R} -  \frac{p - 1}{2} c B'(r) - \frac{p(p - 1)}{4} (c)^{2}}  \nonumber\\
& & + B''(r) - \frac{(B'(r))^{2}}{2} - 2 \Lambda + 2 \kappa_{D} ^{2} t_{0} = 0,
\end{eqnarray}
\begin{equation}
t'_{r} = \frac{p}{2}c (t_{r}(r) - t_{0}(r)) + \frac{1}{2} B'(r)(t_{r}(r) - t_{\theta}(r)).
\end{equation}

From these equations we can get the general solution for the metric:
\begin{equation}
ds^{2} = e^{-cr}\hat{g}_{\mu\nu}dx^{\mu}dx^{\nu}+dr^{2}+ e^{-B(r)}d\theta^{2},
\end{equation}
with
\begin{equation}
B(r) = cr + \frac{4}{pc} \kappa_{D} ^{2} \int{dr (t_{r} - t_{\theta})},
\end{equation}
\begin{equation}
c^{2} = \frac{1}{p(p + 1)} (-8 \Lambda + 8\kappa_{D} ^{2} \alpha)
\end{equation}
and
\begin{equation}
t_{\theta} = \beta e^{cr} + \alpha,
\end{equation}
where $\alpha$ and $\beta$ are constants and $\alpha$ has to satisfy the inequality $-8 \Lambda + 8\kappa_{D} ^{2} \alpha > 0$.

This is a general result. It is possible to derive two special cases from the general  solution above. One of them, the global string-like defect, occurs when the spontaneous symmetry breakdown, $t_{r} = -t_{\theta}$ is present, namely
\begin{equation}\label{metric1}
ds^{2} = e^{-cr}\hat{g}_{\mu\nu}dx^{\mu}dx^{\nu}+dr^{2}+ R_{0} ^{2} e^{-c_{1}r}d\theta^{2},
\end{equation}
where $R_{0} ^{2}$ is a constant. In this case we have
\begin{equation}
c_{1} = c - \frac{8}{pc}\kappa_{D} ^{2} t_{\theta}
\end{equation}
and
\begin{equation}
c^{2} = \frac{1}{p(p + 1)} (-8 \Lambda + 8\kappa_{D} ^{2} \alpha) > 0.
\end{equation}
In the case  $t_{\theta} = t_{r} = 0$ we found a more simplified solution
\begin{equation}\label{metric2}
ds^{2} = e^{-cr}\hat{g}_{\mu\nu}dx^{\mu}dx^{\nu}+dr^{2}+ R_{0} ^{2} e^{-cr}d\theta^{2}
\end{equation}
with
\begin{equation}
c^{2} = \frac{-8 \Lambda}{p(p + 1)}.
\end{equation}
This solution, that can be found setting $t_{i} (r) = 0$, was first showed in Refs. \cite{Gherghetta2000}, and \cite{Gregory2000a}.
Note that the general solution (\ref{eq1}) and the special cases (\ref{cosmo}) and (\ref{eq5}) represent a 4-brane embedded in a six dimensional space-time.
The on-brane dimension $\theta$, $0 \leq \theta \leq 2 \pi$, is compact and assumed to be sufficiently small to realize the 3-brane world.
The other extra dimension $r$ extends to the infinite, $0 \leq r \leq \infty$.

In this same context Koley and Kar \cite{Koley2007} encountered similar results but they followed a different approach. They solve the Einstein equations for two different bulk scalar fields, namely, a phantom field and the Brans-Dicke scalar field. For the first, the phantom field one, the solution represents a 4-brane and for the Brans-Dicke field they encountered a 3-brane in a six dimensional space-time as solution.

Before to close this section it is necessary to explain the difference between global and local string. By local string-like defect we mean the situation
where the energy-momentum tensor is exponentially decreasing or zero outside the core of the thick or thin brane, respectively. This situation is obtained if
we choose $t_{i} = 0$ in (\ref{tensor}). In the case of a global defect, on the other hand, the energy-momentum tensor has a contribution outside the defect. For further considerations and discussions on this subject we suggest the PhD thesis of Roessl \cite{Roessl2004} and the instructive book by Vilenkin and Shellard \cite{VILENKIN1994}.

\section{Localization of the Kalb-Ramond field}

In this section we study the mechanism of field localization for the Kalb-Ramond field in the background geometries defined by eq. (\ref{metric1}) and eq. (\ref{metric2}). We analyze first the zero mode and massive modes later. For the the zero mode we will consider the most general background (\ref{metric1}) while the massive modes will be studied on the local-defect given by (\ref{metric2}).

Before to carry out the localization it is necessary to remember that the Kalb-Ramond field, as a 2-form field, is self-dual in 6D geometry. More than this, it is not so easy to find a Lagrangian formulation for this model which is manifestly Lorentz invariant (MLI). In fact non covariant action for this model were formulated, for example, in the references \cite{Zwanziger1971, Deser1976, Henneaux1987, Schwarz1994}. The MLI model was first performed by Pasti, Sorokin and Tonin, which is called today the PST formalism \cite{Pasti1997}. Before them other MLI models were constructed by McClain \cite {McClain} with an infinite set of auxiliary fields and by Pasti \cite{Pasti1995} with a finite set of auxiliary fields. But in the PST formalism the authors showed that the two last formalisms are equivalents and in fact we need only one auxiliary scalar field to obtain a MLI model for the 2-form chiral boson in 6D geometry.

The action for the MLI chiral 2-form model in PST formalism is given by \cite{Pasti1997}

\begin{equation} \label{PST action}
S =  \int{d}^{6}x \left[-\frac{1}{6}H_{LMN}H^{LMN} + \frac{1}{\partial_{Q} a \partial^{Q} a} \partial^{M} a(x) {\mathbb{H}}_{MNL}{\mathbb{H}}^{NLR} \partial_{R} a(x) \right],
\end{equation}
where $\mathbb{H}_{MNL}$ is an anti self-dual field defined as ${\mathbb{H}}_{MNL} = H_{MNL} - *H_{MNL}$ and $a(x)$ is a scalar field which transforms as a Goldstone field $(\delta a(x) = \varphi(x))$. Since this field can be gauged away it is an auxiliary field. It is important to say that the variation of the action (\ref{PST action}) with respect to $a(x)$ does not produce any additional field equation. In fact it is possible to define a unit time-like vector $u_{M} = \partial_{M} a(x) = \delta_{M} ^{0}$ which will give us another action without the presence of the field $a(x)$ but this model will not be MLI \cite{Pasti1997, medina1997}. Another possibility is to define the space-like vector $\partial_{M} a(x) = \delta_{M} ^{5}$. In this case the action (\ref{PST action}) will take the following form  \cite{Pasti1997, medina1997}
\begin{equation} \label{PST action2}
S =  \int{d}^{6}x \left[-\frac{1}{6}H_{LMN}H^{LMN} + \frac{1}{2} {\mathbb{H}}_{MN5}{\mathbb{H}}^{MN5}\right],
\end{equation}
We see that in this case the auxiliary field $a(x)$ is not present in the action but this is not a MLI model as well. As a matter of fact, it represents the free chiral field formulation given by \cite{Perry}. Finally, it is not possible to put  $u_{M} = \partial_{M} a(x) = 0$ because the $u_{M}$ norm is presented in the denominator of (\ref{PST action}). However, in principle, one can arrange a suitable limit $u_{M} \rightarrow 0$ in a way that preserves the physical contents of the model \cite{Pasti1997}.

By the revision given above about the PST formalism we see that in order to construct a MLI model for the Kalb-Ramond field in 6D geometry, it is necessary to consider that this field is self-dual. On the other hand, in order to implement self duality in the model it is necessary to use (at least) an auxiliary field. However it is possible to construct interesting model, like the free chiral field formulation, without the necessity of manifested Lorentz invariance.

So, in this work we will not focus on the self duality nature of the KR field. We will show that even in this case we can have KR zero mode
localization on the brane. We argue that this is because we are in a warped geometry, given by eq. (\ref{metric1}) or eq. (\ref{metric2}), on the contrary
that happens in PST formalism, which is a Minkowski 6D geometry. Therefore it is possible to localize field using the gravitational interaction
only without the necessity of an auxiliary field.

\subsection{The zero-mode case}

In this subsection we will show that, for the Kalb-Ramond field,  the zero mode is localized on a string-like defect since the constants $c$ and $t_{\theta}$ could obey some specific inequality conditions.

From the action of the Kalb-Ramond tensor field
\begin{equation}
S_{m} = \frac{-1}{12} \int{d}^{D}x \sqrt{-g}g^{MQ}g^{NR}g^{LS}H_{MNL}H_{QRS} \label{KR},
\end{equation}
we derive the equation of motion
\begin{equation}
\partial_{Q}[ \sqrt{-g}H_{MNL}g^{MQ}g^{NR}g^{LS}] = 0.
\end{equation}
A straightforward calculation give us the following form for the equation of motion
\begin{equation}
P^{-1}(r)\partial_{\mu}H^{\mu\sigma\beta} + P^{-p/2 + 2}(r)Q^{-1/2}(r)\partial_{r}[P^{p/2 - 2}(r)Q^{1/2}(r)H_{r} ^{\sigma\beta}] + Q^{-1}(r)\partial_{\theta}H_{\theta} ^{\sigma\beta} = 0,
\end{equation}
where we redefine $\hat{g}_{\mu\nu} = \eta_{\mu\nu}$, $e^{-A(r)} = P(r)$ and $R_{0} ^{2}e^{-B(r)} = Q(r)$. Although we know that $A(r) = cr$ and $B(r) = c_{1}r$, we keep the  forms $A(r)$ and $B(r)$ in order to be more general.

Let us assume the gauge condition $B^{\mu r} = B^{\mu \theta} = 0$ and the decomposition
\begin{equation}
B^{\mu \nu}(x^{M}) = b^{\mu\nu}(x^{\mu}) \sum{\rho_{m}(r) e^{il\theta}}
\end{equation}
\begin{equation}
B^{r \theta}(x^{M}) = b^{r \theta}(x^{\mu}) \sum{\rho_{m}(r) e^{il\theta}}.
\end{equation}
By choosing $\partial_{\mu} h^{\mu\sigma\beta} = m_{0} ^{2} b^{\sigma\beta}$ we get the following equation for the radial variable
\begin{equation}\label{gran}
\partial_{r} ^{2} \rho_{m}(r) + \left( (\frac{p}{2} - 2)\frac {P^{'}(r)}{P(r)} + \frac{Q^{'}(r)}{2Q(r)}\right) \partial_{r} \rho_{m}(r) + \left( \frac{1}{P(r)} m_{0} ^{2} - \frac{1}{Q(r)} l^{2} \right) \rho_{m}(r) = 0,
\end{equation}
where $l$ is the angular quantum number.

This equation has the zero mode $(m_{0} = 0)$ and s-wave $(l = 0)$ and $b^{r \theta}$ is a $constant$. Note that $b^{r \theta}$ need not to be constant. This is a special case in order to get localization similar to the ones in scalar and vector field solutions \cite{Oda2000a}.

If we substitute the constant solution in
the action (\ref{KR}), and remember the forms of $A(r)$ and $B(r)$, than the $r$ integral reads
\begin{equation}\label{action1}
I_{0} \propto \int{dr P^{p/2 - 3} Q^{1/2}} \propto \int{dr e^{-[(p/2 - 3)c + 1/2c_{1}]r}}.
\end{equation}
But we need $I_{0}$ to be finite. It is possible to rewrite this condition as an inequality, namely, for $c > 0$,
\begin{equation}\label{cond1}
\frac{1}{\kappa_{D} ^{2}} \Lambda < t_{\theta} < \frac{-(p - 5)}{ 6 \kappa_{D} ^{2}} \Lambda
\end{equation}
and
\begin{equation}\label{cond2}
t_{\theta} > \frac{-(p - 5)}{ 6 \kappa_{D} ^{2}} \Lambda
\end{equation}
for $c<0$.

The conditions (\ref{cond1}) and (\ref{cond2}) are very similar to the ones encountered for the scalar and vector fields \cite{Oda2000}. Note, however, that if we set $p = 4$ in (\ref{action1}) and additionally set $c_{1} = c$, which is nothing but the string-like defect for the metric (\ref{metric2}), we need to have $c < 0$ in order to have localized KR zero mode. Therefore the scalar and vector field zero modes in this context are localized only if $c > 0$.

\subsection{The massive modes case}

Now we are going to analyze the massive modes. Although we will only study localization in the background geometry of the metric (\ref{metric2}), we will begin by
considering the general background of the metric (\ref{metric1}).

So we now return to equation (\ref{gran}) and make the following changes of variable $r\Rightarrow z = f(r)$ and $\rho_{m} (r) = \Omega (r)\Psi (r)$ with
\begin{equation}\label{rel}
\frac{d z}{d r} = e^{A(r)/2}
\end{equation}
and
\begin{equation}
\Omega (r) = e^{- \left( \frac{ (5 - p) A(r)}{4} - \frac{B(r)}{4}\right)}.
\end{equation}
This give us the following Schrodinger-like equation
\begin{equation}\label{schro}
\left[ - \frac{d^{2}}{d z^{2}} + V(z) \right] \Psi = m_{n} ^{2} \Psi,
\end{equation}
where
\begin{equation}\label{pot}
V(z) = \frac{1}{2} \ddot{C(z)} + \frac{1}{4} \dot{C}^{2} (z) + \frac{l^{2}}{R_{0} ^{2}} e^{B(z) - A(z)}
\end{equation}
with
\begin{equation}\label{C1}
\dot{C}(z) = \left[ \frac{ (5 - p) \dot{A}(z)}{2} - \frac{ \dot{B}(z)}{2}\right]
\end{equation}
and
\begin{equation}\label{C2}
\ddot{C}(z) = \left[ \frac{ (5 - p) \ddot{A}(z)}{2} - \frac{ \ddot{B}(z)}{2}\right],
\end{equation}
where dot represents differentiation with respect to the variable $z$. It is still possible to rewrite eq. (\ref{schro}) in the supersymmetric quantum mechanics scenario
\begin{equation}\label{sqm}
Q^{\dagger} Q\Psi = \left\{ -\frac{d}{dz} - \frac{1}{2} \dot{C} + \frac{l}{m_{n} } e^{\frac{1}{2}(B - A)}\right\} \left\{ \frac{d}{dz} - \frac{1}{2} \dot{C} + \frac{l}{m_{n} } e^{\frac{1}{2}(B - A)}\right\} \Psi = m_{n} ^{2} \Psi,
\end{equation}
which holds only for $p = 4$. Nevertheless, if we set $l = 0$ it is ever possible to rewrite eq. (\ref{schro}) in the form given by eq. (\ref{sqm}).

Now we explicitly write the potential (\ref{pot}). Keeping in mind the relation (\ref{rel}), we find that
\begin{equation}\label{pot1}
V(z) = (-C_{2} + C_{2} ^{2})\frac{1}{z^{2}} + \frac{l^{2}}{R_{0} ^{2}}\left( \frac{z^{2} c^{2}}{4}\right) ^{(\frac{c_{1}}{c} - 1)},
\end{equation}
where $C_{2}$, which depends only on the constants $p$, $c_{1}$ and $c$, is given by
\begin{equation}\label{const}
C_{2} = \frac{1}{2}(5 - p - \frac{c_{1}}{c}).
\end{equation}
It is still possible to write the potential in terms of the $r$ variable. In this case we have
\begin{equation}\label{pot2}
V(r) = e^{-cr}\left[\frac{1}{4}C'^{2}(r) + \frac{1}{4}D'(r)\right] + \frac{l^{2}}{R_{0} ^{2}} e^{c_{1} - c},
\end{equation}
where
\begin{equation}
C'(r) = \left[ \frac{ (5 - p) A'(r)}{2} - \frac{ B'(r)}{2}\right]
\end{equation}
and
\begin{equation}
D'(r) = \left[ \frac{ -(5 - p) A'^{2}(r)}{2} - \frac{ B'(r)A'(r)}{2}\right].
\end{equation}
Remembering that prime represents differentiation with respect to $r$. We remark that in the explicit forms of the potentials (\ref{pot1}) and (\ref{pot2}) we have used the fact that $A(r) = cr$ and $B(r) = c_{1}r$ and also that $z = (2/c) e^{cr/2}$, obtained from eq. (\ref{rel}).

Now that we have the Schrodinger-type equation (\ref{schro}) and the general potential (\ref{pot1}) we can analyze and discuss about the localization properties of the massive modes for the Kalb-Ramond field. We will only consider the local string-like defect $c_{1} = c$. So applying this condition in equations (\ref{pot1}) and (\ref{const}) that define, respectively, the potential $V(z)$ and the constant $C_{2}$ and making $p = 4$, the Schrodinger-type equation (\ref{schro})
takes the following simple form
\begin{equation}\label{schro1}
\left[ \frac{d^{2}}{d z^{2}} + M_{n} ^{2} \right] \Psi = 0
\end{equation}
where $M_{n} ^{2} = m_{n} ^{2} - \frac{l^{2}}{R_{0} ^{2}}$. The solution to this equation is simply
\begin{equation}
\Psi(z) = \alpha_{1} Sen\left[ \sqrt{\left( m_{n} ^{2} - \frac{l^{2}}{R_{0} ^{2}}\right)}z\right] + \alpha_{2} Cos\left[ \sqrt{\left( m_{n} ^{2} - \frac{l^{2}}{R_{0} ^{2}}\right)}z\right]
\end{equation}
where $\alpha_{1}$ and $\alpha_{2}$ are some integration constants. In terms of the $r$ variable we have the solution

\begin{equation}
\Psi(r) = \alpha_{1} Sen\left[ \sqrt{\left( m_{n} ^{2} - \frac{l^{2}}{R_{0} ^{2}}\right)}(2/c) e^{cr/2}\right] + \alpha_{2} Cos\left[ \sqrt{\left( m_{n} ^{2} - \frac{l^{2}}{R_{0} ^{2}}\right)}(2/c) e^{cr/2}\right]
\end{equation}
By the fact that the operator in eq. (\ref{schro1}) is self-adjoint we can impose the boundary conditions
\begin{equation}\label{cond3}
\Psi'(0) = \Psi'(\infty) = 0
\end{equation}
The condition (\ref{cond3}), for $l=0$, implies $\alpha_{1} = 0$. So the general solution for the eq. (\ref{schro}) assumes the form
\begin{equation}\label{psi}
\Psi(z) =  \alpha_{2} Cos\left[ \sqrt{\left( m_{n} ^{2} - \frac{l^{2}}{R_{0} ^{2}}\right)}z\right].
\end{equation}

\begin{figure}[!htb]
\begin{minipage}[b]{0.45\linewidth}
\includegraphics[width=5cm]{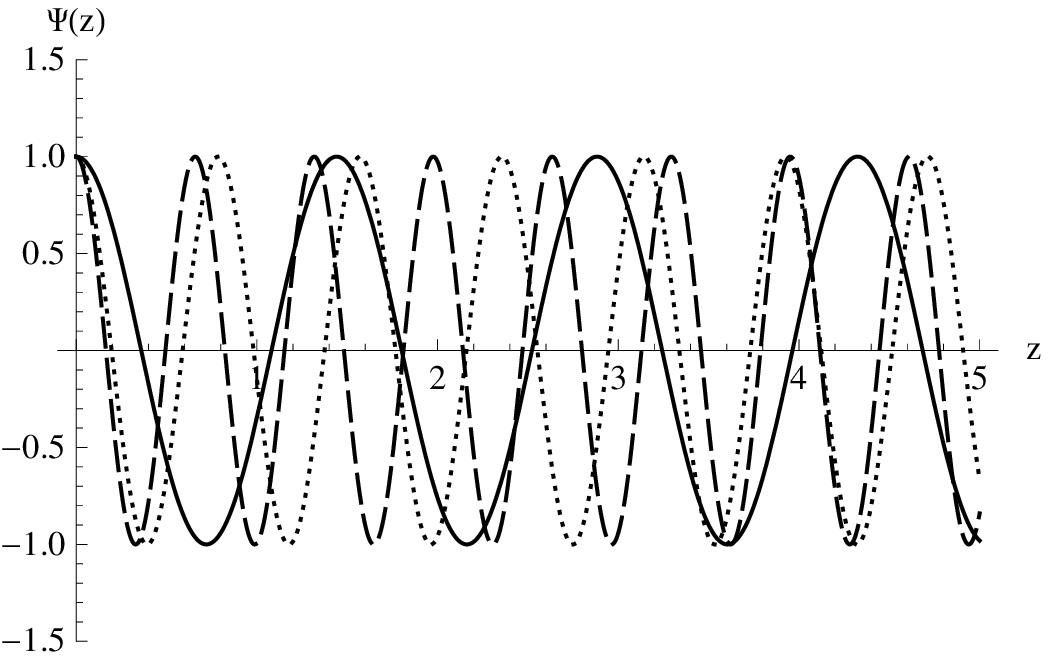}
\caption{\label{fig:onda} Wave function for $l=3$ (dashed line), $l=6$ (dotted line) and $l=9$ (solid line). We keep $R_{0} = 1$ and $m = 10$.}
\end{minipage}\hfill
\begin{minipage}[b]{0.45\linewidth}
\includegraphics[width=5cm]{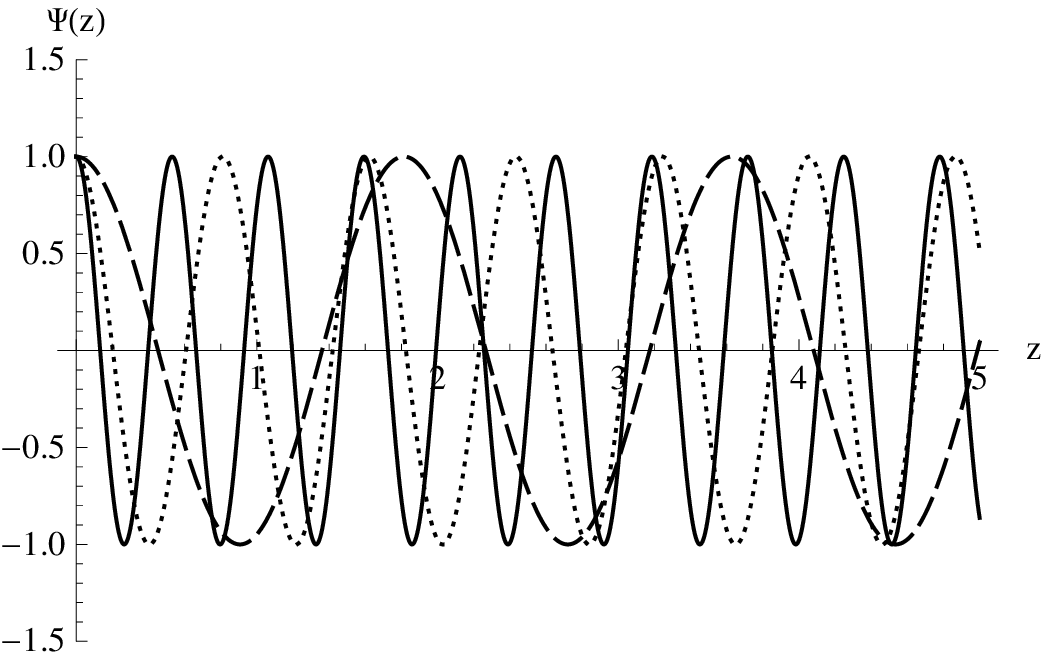}
\caption{\label{fig:onda2} Wave function for $l=2$ and $R_{0} = 1$ and distinct mass values $m=4$ (dashed line), $m=8$ (dotted line) and $m=12$ (solid line).}
\end{minipage}
\end{figure}

For $\alpha_{2} = 1$ we plot the function (\ref{psi}) in two different situations: in Fig. (\ref{fig:onda}) we keep $R_{0} = 1$ and $m = 10$ as constants and we take distinct values of $l$. In Fig. (\ref{fig:onda2}) $R_{0} = 1$ and $l = 2$ are held constants while we take different values for $m$. As can be seen, the variation in the angular quantum number induces phase and frequency wave changes, but do not affect the wave amplitude. However, while the frequency increases with the mass increasing, the contrary occurs with respect to the angular quantum number, since for increasing $l$ we have decreasing wave frequency.

Considering the analysis of massive modes in some previous works \cite{csaba,gremm,nosso2,nosso3} we note that, for some values of mass, our plane
 wave solutions of the Schrodinger equation can assume very high amplitudes inside the brane and this may be understood in terms of resonance structures. Analyzing
 the figures (\ref{fig:onda}) and (\ref{fig:onda2}) we observe only variations in the period of the oscillations without high amplitudes in $z=0$.  These two modes
do not present resonances. However, a better way to detect resonant structures consists in evaluate the value of our solutions inside the defect as function of the mass.
This method has been used successfully to detect resonant modes in numerical solutions of the Schrodinger equation \cite{nosso2,nosso3,ca,ca-wi}. However
the type of solutions that we have obtained from the Schrodinger-like equations are very simple here in order to consider this numerical method. From solutions (\ref{psi}) we can conclude
that there is no resonances in the present scenario.

\subsection{ Kaluza-Klein decomposition}

Now we are going to analyze the massive modes for the Kalb-Ramond field, via the so called Kaluza-Klein decomposition. We begin by rewriting the equation (\ref{gran}), with $Q(r) = R_{0} ^{2} P(r)$, in the following form

\begin{equation} \label{49}
-\left[ P^{-\frac{p - 5}{2}}\partial_{r} \left( P^{\frac{p - 3}{2}}\partial_{r}\rho_{m}\right) - \frac{l^{2}}{R_{0} ^{2}}\right]\rho_{m} = m_{n} ^{2}\rho_{m}.
\end{equation}

Note that the operator above is self-adjoint, so we can impose the orthonormality condition

\begin{equation}
\int_{0} ^{\infty}{dr P^{\frac{p - 5}{2}}\rho_{m}\rho_{m'}} = \delta_{mm'}.
\end{equation}

Now we perform the change of variables $z = 2/c M_{n} P^{-1/2}$ and $\rho = P^{-(p - 3)/4} h$ in the eq. (\ref{49}). This give us the following Bessel equation of order $(p - 3)/2$
\begin{equation}
\frac{d^{2}h}{dz^{2}} + \frac{1}{z}\frac{d h}{dz} + \left[ 1 - \frac{1}{z^{2}}\left( \frac{p - 3}{2}\right)^{2} \right]h = 0. \label{bessel}
\end{equation}
The solution of this equation is straightforward, so the radial function $ \rho $ is given by
\begin{equation}		
\rho(z) =  \frac{1}{N_{n}}P^{-(p - 3)/4} \left[  J_{(p - 3)/2} (z) + \alpha_{n} Y_{(p - 3)/2} (z) \right],
\end{equation}
where $N_{n}$ are normalization constants and $\alpha_{n}$ are constant coefficients.

Provided that the operator in (\ref{bessel}) is self-adjoint, we can now impose the boundary conditions below,
\begin{equation} \label{53}
\rho'(0) = \rho'(\infty) = 0.
\end{equation}
These boundary conditions give us the constant coefficients $\alpha_{n}$
\begin{equation}
\alpha_{n} = -\frac{J_{\frac{p-5}{2}(z_{n}(0))}}{Y_{\frac{p-5}{2}(z_{n}(0))}} = -\frac{J_{\frac{p-5}{2}(z_{n}(\overline{r}))}}{Y_{\frac{p-5}{2}(z_{n}(\overline{r}))}}.
\end{equation}
For $M_{n} << c$ the KK masses can be derived \cite{Oda2000a} from the formula
\begin{equation} \label{55}
J_{(p - 5)/2}(z(\overline{r})) = 0,
\end{equation}
where $\overline{r}$ is the infrared cutoff which will be extended to infinity in the end of calculations. Finally using (\ref{53}) and (\ref{55}) we get the approximate mass formula
\begin{equation} \label{56}
M_{n} = \pi \frac{c}{2} \left[ n + \frac{p}{4} - \frac{3}{2}\right]e^{-c\overline{r} /2}
\end{equation}
and the approximate formula for the normalization constants
\begin{equation}
N_{n} = \sqrt{c}\frac{z_{n}(\overline{r})}{2M_{n}}J_{\frac{p-3}{2}(z_{n}(\overline{r}))}.
\end{equation}
From (\ref{56}) we see that in the limit $\overline{r} \rightarrow \infty$, the KK masses of the Kalb-Ramond field depend on $l^{2}/R_{o} ^{2}$, so the only massless mode is the s-wave $l = 0$, and consequently the other ones are massive. These results are in accordance with the ones found for the gravity field by Gherghetta and Shaposhnikov \cite{Gherghetta2000} and for the scalar and vector fields by Oda \cite{Oda2000a}. Therefore we are equipped with a model which give us desirable physical properties.

\section{Discussions}

In this work we have studied the Kalb-Ramond tensor field in the bulk in a string like defect model. This model was studied
by other authors  \cite{Oda2000, Gregory2000, Gherghetta2000a} and we summarized its features in the second section of
this work. It is a Randall-Sundrum type model \cite{Randall1999a,Randall1999} in six dimensions and represents
a 4-brane embedded in a six dimensional space-time. One of the extra dimensions is compact, $\theta$, $0 \leq \theta \leq 2 \pi$,
an the other one is semi infinite, $0 \leq r \leq \infty$. The bulk is an $AdS_{6}$ space-time with positive cosmological constant.
Starting by the action for the KR field we derived the equation of motion and study the possibility to have zero mode localization on the brane and additionally we study the massive modes in two different ways, namely, by analyzing massive modes and by calculating the Kaluza-Klein mass spectra.

Unlike the case studied in five dimensions where the Kalb-Ramond field is not localized only by gravitational interactions \cite{Tahim2009}, in the present scenario we have zero mode localization. The main advantage of this scenario \cite{Oda2000,Oda2000a} over the membrane models is the possibility to localize fields of various spins without include other fields in the bulk. As noted in Ref. \cite{Tahim2009} a necessary condition to obtain localization of zero mode tensor gauge field in membrane models is the inclusion of the dilaton field to the background. This condition was not necessary in this work. While for the scalar and gauge fields the localization, in the local defect, occurs for an exponentially decreasing warp factor \cite{Oda2000}, in the KR field case it is necessary to have an exponentially increasing warp factor, as in the fermionic field localization \cite{Oda2000a, Gherghetta2000}, in order to obtain zero mode localization.

In order to analyze the massive modes we write the equations of motion for the fields in the form of the Schrodinger equation avoiding tachyonic states. While in membrane  scenarios in five dimensions the analysis of massive modes is possible only numerically \cite{gremm,nosso2,nosso3,ca}, in our case we obtain analytical solutions for the massive modes. Usually, the massive modes analysis is complemented by the search of resonant modes in the spectrum, however the form of our solutions to local defects are sufficient to exclude the existence of resonant modes. So in this case it is not interesting to study the massive modes by resonant modes procedure. Nevertheless, for the global string case this method may be
interesting.

Other way to analyze the massive modes in the context of brane worlds in six dimensions, is to perform the
so called Kaluza-Klein decomposition. This means to solve the equation for the radial extra dimension and evaluate the mass
spectra from it. In our case, the approximate mass spectra is similar to the one encountered for the scalar and
gauge fields \cite{Oda2000a}  differing from them only by  $-1$ and $-1/2$, respectively, in the quantity between
parenthesis in equation (\ref{56}). So in this point we complement the works of Oda \cite{Oda2000a}, which did the
same calculation for the scalar and vector fields, and Gherghetta \cite{Gherghetta2000}, which studied
the spin 2 graviton field.

Our next step is to look for the extension of this work to the global defect case, particularly, the calculations of the mass spectra. It is important to say that in this context there is no similar work, so it is interesting to perform this work, not only to the Kalb-Ramond field but for the other Standard Model fields.

The authors would like to thank FUNCAP, CNPq and CAPES (Brazilian agencies) for financial support.


\begin{thebibliography}{99}
\bibitem{Rubakov1983} V.A. Rubakov, M.E. Shaposhnikov, Phys. Lett. \textbf{B125}, 139 (1983).
\bibitem{YuriShtanov2009} Y. Shtanov, V. Sahni, A. Shafieloo and A. Toporensky, JCAP \textbf{04}, 023 (2009).
\bibitem{G.D.Starkman2001a} G. D. Starkman, D. Stojkovic and M. Trodden, Phys. Rev. Lett. \textbf{87}, 231303 (2001).
\bibitem{G.D.Starkman2001} G. D. Starkman, D. Stojkovic and M. Trodden, Phys.Rev. \textbf{D63}, 103511 (2001).
\bibitem{I.Antoniadis1998} I. Antoniadis, N. Arkani-Hamed and S. Dimopoulos, G. Dvali, Phys. Lett. \textbf{B436}, 257 (1998).
\bibitem{N.Arkani-Hamed2000} N. Arkani-Hamed, S. Dimopoulos and G. Dvali, Phys.Lett. \textbf{B429}, 263 (1998).
\bibitem{Randall1999a} L. Randall and R. Sundrum, Phys. Rev. Lett. \textbf{83}, 3370 (1999).
\bibitem{Randall1999} L. Randall and R. Sundrum, Phys. Rev. Lett. \textbf{83},  4690 (1999).
\bibitem{MukhopadhyayaB2002} B. Mukhopadhyaya, S. Sen, S. SenGupta, Phys. Rev. Lett. \textbf{89}, 121101 (2002).
\bibitem{Mukhopadhyaya2004} B. Mukhopadhyaya, Siddhartha Sen, Somasri Sen, S. SenGupta, Phys. Rev. \textbf{D70}, 066009 (2004).
\bibitem{Lebedev2002} O. Lebedev, Phys. Rev. \textbf{D65}, 124008 (2002).
\bibitem{Oda2000} I. Oda, Phys. Lett. \textbf{B496}, 113 (2000).
\bibitem{Davoudias2000} H. Davoudiasl, J.L. Hewett, T.G. Rizzo, Phys. Rev. \textbf{D63}, 075004 (2001).
\bibitem{Kehagias2001} A. Kehagias, K. Tamvakis, Phys. Lett. \textbf{B504}, 38 (2001).
\bibitem{Tahim2009} M. O. Tahim, W. T. Cruz and C. A. S. Almeida, Phys. Rev. \textbf{D79}, 085022 (2009).
\bibitem{Oda2000a} I. Oda, Phys. Rev. \textbf{D62}, 126009 (2000).
\bibitem{Gherghetta2000} T. Gherghetta and M. Shaposhnikov, Phys. Rev. Lett. \textbf{85} 240 (2000).
\bibitem{Akama2000} K. Akama and T. Hattori, Mod. Phys. Lett. \textbf{A15} 2017 (2000).
\bibitem{Visser1985} M. Visser, Phys. Lett. \textbf{B159}, 22 (1985).
\bibitem{Gregory2000a} R. Gregory, V.A. Rubakov and S. M. Sibiryakov,  Phys. Rev. Lett. \textbf{84}, 5928 (2000).
\bibitem{Koley2007} R. Koley and S. Kar, Class. Quant. Grav. \textbf{24}, 79 (2007).
\bibitem{Roessl2004} E. E. Roessl, Topological defects and gravity in theories with extra dimensions, Phd Thesis. hep-th/0508099 (2004)
\bibitem{VILENKIN1994} A. Vilenkin and E. P. S. Shellard, Cosmic strings and other topological defects, Cambridge University Press. (1994).
\bibitem{Zwanziger1971} D. Zwanziger, Phys. Rev. \textbf{D3}, 880 (1971).
\bibitem{Deser1976} S. Deser and C. Teitelboim, Phys. Rev. D13, 1592 (1976); S. Deser, J. Phys. A: Math. Gen. \textbf{15}, 1053 (1982).
\bibitem{Henneaux1987} M. Henneaux and C. Teitelboim, in Proc. Quantum mechanics of fundamental systems 2, Santiago, 1987, p. 79; Phys. Lett. \textbf{B206}, 650 (1988).
\bibitem{Schwarz1994} J. H. Schwarz and A. Sen, Nucl. Phys. \textbf{B411}, 35 (1994).
\bibitem{Pasti1997} P. Pasti, D. Sorokin and M. Tonin. Lorentz-invariant actions for chiral $p$-forms. Phys. Rev. \textbf{D55}, 6292 (1997).
\bibitem{McClain} B. McClain, Y. S. Wu, F. Yu, Nucl. Phys. \textbf{B343}, 689 (1990).
\bibitem{Pasti1995}P. Pasti, D. Sorokin and M. Tonin, Phys. Lett. \textbf{B352}, 59 (1995); P. Pasti, D. Sorokin and M. Tonin, Phys. Rev. \textbf{D52}, R4277 (1995); P. Pasti, D. Sorokin and M. Tonin, in Leuven Notes in Mathematical and Theoretical Physics, (Leuven University Press) Series B V6, 167 (1996), hep–th/9509053.
\bibitem{medina1997} R. Medina, N. Berkovits, Pasti-Sorokin-Tonin actions in the presence of sources. Phys. Rev. \textbf{D56}, 6388 (1997).
\bibitem{Perry} M. Perry and J. H. Schwarz, Interacting chiral gauge fields in six dimensions and Born-Infeld theory, Nucl. Phys. \textbf{B489}, 47 (1997), hepth/9611065.
\bibitem{csaba}C. Csaki, J. Erlich, T. J. Hollowood, Y. Shirman, Nucl. Phys. \textbf{B581}, 309 (2000).
\bibitem{gremm} M. Gremm, Phys. Lett. \textbf{B478}, 434 (2000).
\bibitem{nosso2} W.T. Cruz, M.O. Tahim and C. A. S. Almeida, Phys Lett. \textbf{B686}, 259-263 (2010).
\bibitem{nosso3} W.T. Cruz, M.O. Tahim and C. A. S. Almeida, Europhys. Lett. \textbf{88}, 41001 (2009).
\bibitem{ca} C. A. S. Almeida, M. M. Ferreira, Jr., A. R. Gomes, R. Casana, Phys. Rev. \textbf{D79}, 125022 (2009).
\bibitem{ca-wi} W.T. Cruz and C. A. S. Almeida, Eur. Phys. J. \textbf{C71}, 1709 (2011).


\end{thebibliography}
\end{document}